# Is big team research fair in national research assessments? The case of the UK Research Excellence Framework 2021

Mike Thelwall
Statistical Cybermetrics and Research Evaluation Group, University of Wolverhampton, UK.
https://orcid.org/0000-0001-6065-205X m.thelwall@wlv.ac.uk

Kayvan Kousha
Statistical Cybermetrics and Research Evaluation Group, University of Wolverhampton, UK.
https://orcid.org/0000-0003-4827-971X k.kousha@wlv.ac.uk

Meiko Makita
Statistical Cybermetrics and Research Evaluation Group, University of Wolverhampton, UK.
https://orcid.org/0000-0002-2284-0161 meikomakita@wlv.ac.uk

Mahshid Abdoli
Statistical Cybermetrics and Research Evaluation Group, University of Wolverhampton, UK.
https://orcid.org/0000-0001-9251-5391 m.abdoli@wlv.ac.uk

Emma Stuart
Statistical Cybermetrics and Research Evaluation Group, University of Wolverhampton, UK.
https://orcid.org/0000-0003-4807-7659 emma.stuart@wlv.ac.uk

Paul Wilson
Statistical Cybermetrics and Research Evaluation Group, University of Wolverhampton, UK.
https://orcid.org/0000-0002-1265-543X pauljwilson@wlv.ac.uk

Jonathan Levitt
Statistical Cybermetrics and Research Evaluation Group, University of Wolverhampton, UK.
https://orcid.org/0000-0002-4386-3813 j.m.levitt@wlv.ac.uk

Collaborative research causes problems for research assessments because of the difficulty in fairly crediting its authors. Whilst splitting the rewards for an article amongst its authors has the greatest surface-level fairness, many important evaluations assign full credit to each author, irrespective of team size. The underlying rationales for this are labour reduction and the need to incentivise collaborative work because it is necessary to solve many important societal problems. This article assesses whether full counting changes results compared to fractional counting in the case of the UK's Research Excellence Framework (REF) 2021. For this assessment, fractional counting reduces the number of journal articles to as little as 10% of the full counting value, depending on the Unit of Assessment (UoA). Despite this large difference, allocating an overall grade point average (GPA) based on full counting or fractional counting give results with a median Pearson correlation within UoAs of 0.98. The largest changes are for Archaeology (r=0.84) and Physics (r=0.88). There is a weak tendency for higher scoring institutions to lose from fractional counting, with the loss being statistically significant in 5 of the 34 UoAs. Thus, whilst the apparent over-weighting of contributions to collaboratively authored outputs does not seem too problematic from a fairness perspective overall, it may be worth examining in the few UoAs in which it makes the most difference.

**Keywords**: Collaboration; Research assessment; REF; REF2021; Research quality; Scientometrics.

# 1 Introduction

When an academic produces collaborative outputs, they create a problem for those that need to evaluate their work, including current and future employers, and national research evaluation systems. Even if all authors of an output complete a CRediT (casrai.org/credit/) statement, the quality of each author's contribution will not be clear, and neither will the exact percentage of the credit that should be assigned to them. For example, should a job candidate be given 1/n of the credit for their articles with n authors, in the absence of any other information about their share (fractional counting)? Or should first authors be assigned 100% of the credit and the others none (first author counting, also known as straight counting)? Or should they be assigned full credit for all articles (full counting)? Historically, the choice of counting method has usually been based on how well they represent the contributions of authors, but studies have often been forced to use the simplest method, whole (full) counting, for practical reasons (Gauffriau, 2021).

Although versions of fractional counting are the fairest and most consistent bibliometrically (Waltman, 2016), whole or complete counting has been more used in bibliometric studies (Van Hooydonk, 1997). This can change the results (Gauffriau & Larsen, 2005). Counting methods may be thought to measure participation (full counting), contribution (fractional counting variants) and leadership (first author counting, assuming that the first author led or conducted most of the research, even if a senior last author guided it) (e.g., Moed, 2005). Other proposed variants include harmonic counting in which authors receive decreasing credit fractions starting with the first author (Hagen, 2010), the more complex modified fractional counting (Sivertsen et al., 2019), and at least 28 others (Gauffriau, 2021). Whilst harmonic counting might seem fairer than fractional counting, some fields still use alphabetical ordering, some collaborations have roughly equal contributions and some authorship lists are alphabetical in the middle (Levitt & Thelwall, 2013; Mongeon et al., 2017).

The choice of counting method can make a clear difference even on a national scale (Aksnes et al., 2012; Sivertsen et al., 2019, p. 680) and so it is important to consider the issue carefully for important applications. National Performance-Based Funding Systems (PBFS) (Hicks, 2012) are high profile examples where counting method choice can have substantial policy, financial and reputational impacts. Full counting may be preferred when reputation or funding is at stake in the belief that collaborative research is good and should be incentivised (Bloch & Schneider, 2016) but the influence of the decision should still be assessed. Full counting is used in the BOF-key in Flanders in Belgium (Debackere & Glänzel, 2004; Engels & Guns, 2018), fractional counting in Australia (Woelert & McKenzie, 2018), Norway (Sivertsen, 2018), and South Korea (Jeon & Kim, 2018) and a weighted version of fractional counting in Denmark (Nielsen, 2017). Denmark's weighted version is between standard fractional counting and full counting, presumably to incentivise collaboration. In general, the introduction of performance-based research funding seems to increase collaboration (Bloch & Schneider, 2016; Bouabid, 2014).

The UK Research Excellence Framework (REF) uses a full counting model. Unlike most PBFS, it uses peer review (for quality scores from 1* to 4*) and limits the number of outputs that each researcher can submit: 4 in REF2014 and 5, with an average of 2.5, in REF2021 (REF2021, 2020). Both versions assessed the outputs (e.g., articles, books, chapters, papers, compositions), research environment, and societal impacts of publicly funded UK research institutions, with the results giving both prestige and block grants for research. For REF2021 the research was split into 34 Units of Assessment (UoAs), each encompassing a range of related academic fields and with a team of senior mainly academic experts to make the peer review evaluations. The number of outputs per researcher must be capped to make peer review practical. This makes full counting for the results a natural choice because otherwise researchers would be directly penalised for collaborating, or the assessors would have to review impractically many outputs if fractional credit was used (e.g., 10 times as many in UoAs with usually 10 authors per paper). Nevertheless, this process is clearly unfair, as a theoretical extreme case shows. Suppose that Department S only submits solo-authored research and Department C only submits big team science with 1000 authors. Then, other factors being equal, Department S is equally rewarded with Department C, despite having done 1000 times more work. The Grade Point Average (GPA) for each institution and UoA is usually calculated as the simple average of the scores for the outputs, combined with the environment and impact scores. The GPA is not an official part of the REF and is not displayed with the results, but universities and newspapers calculate this and rank submissions within each UoA. Nevertheless, the GPA could reasonably be calculated with fractional counting even if full counting was used for the funding formula.

Full counting in the REF is apparently not controversial despite its obvious unfairness, but the rules for dealing with collaborative articles are problematic. Whilst a report on research evaluation acknowledged that collaboration was the norm in many areas of science and that this masked the contributions of authors, no practical solution could be found (Wilsdon et al., 2015). It is not clear whether fractional counting for UK REF GPAs would have a substantial influence on the results in terms of the relative scores of universities or their rankings. This issue is investigated here with REF2021 data.

## 2   Methods

Provisional article-level REF2021 results from March 2022 were supplied by the REF team for all journal outputs submitted for evaluation except those from the University of Wolverhampton (for confidentiality reasons). The information included the institution submitting each article, the UoA (out of 34), the provisional score (0 to 4) and the number of authors.

GPAs were calculated separately for each institution within each UoA using whole counting, where each author is credited for the full score for their articles, irrespective of the number of co-authors, and fractionalised weighting (complete-fractionalised in the terminology of Gauffriau, 2021). For this, an institution submitting an article with n authors is credited with 1/n of its score, where n is the total number of authors of the article.

Within each UoA the institutional full weighting GPAs were correlated against the institutional fractional weighting GPAs to assess the extent to which they gave equivalent results. Overall, fractional counting would give lower GPAs because more collaborative articles tend to be higher quality, but since reputations and funding in the REF are essentially zero-sum games, relative scores are more relevant than absolute scores.

# 3   Results

The number of institutions with at least one journal article submitted to REF2021, excluding the University of Wolverhampton, varied from 17 (Classics) to 107 (Business and Management Studies). The number of articles submitted varied from 227 to 17,929, or from 198.4 to 9993.6 if fractional counting is used (Table 1). The large variations are mainly due to differing numbers of outputs submitted to each UoA and the proportion of non-article outputs submitted to each one, such as books, book chapters, and artworks.

Table 1. Sample sizes for journal articles in the 34 REF2021 UoAs.

| Name | Institutions | Articles | Articles (fractional) |
|---|---|---|---|
| 1: Clinical Medicine | 31 | 11972 | 1255.3 |
| 2: Public Health, Health Services & Primary Care | 33 | 4900 | 690.4 |
| 3: Allied Health Professions, Dentistry, Nursing & Pharmacy | 89 | 11441 | 2476.0 |
| 4: Psychology, Psychiatry & Neuroscience | 92 | 9711 | 2347.3 |
| 5: Biological Sciences | 44 | 7098 | 1054.6 |
| 6: Agriculture, Food & Veterinary Sciences | 25 | 3423 | 574.9 |
| 7: Earth Systems & Environmental Sciences | 40 | 4356 | 887.0 |
| 8: Chemistry | 41 | 3688 | 652.5 |
| 9: Physics | 44 | 5482 | 870.3 |
| 10: Mathematical Sciences | 54 | 5819 | 2567.3 |
| 11: Computer Science & Informatics | 89 | 5547 | 1893.4 |
| 12: Engineering | 88 | 17929 | 5209.2 |
| 13: Architecture, Built Environment & Planning | 37 | 2996 | 1339.3 |
| 14: Geography & Environmental Studies | 56 | 4028 | 1537.4 |
| 15: Archaeology | 24 | 693 | 286.1 |
| 16: Economics & Econometrics | 25 | 2128 | 1044.1 |
| 17: Business & Management Studies | 107 | 15562 | 6693.6 |
| 18: Law | 68 | 3385 | 2778.9 |
| 19: Politics & International Studies | 56 | 3065 | 2106.5 |
| 20: Social Work & Social Policy | 75 | 4000 | 2142.6 |
| 21: Sociology | 37 | 1753 | 1166.5 |
| 22: Anthropology & Development Studies | 22 | 1155 | 724.6 |
| 23: Education | 82 | 4073 | 2257.7 |
| 24: Sport & Exercise Sciences, Leisure & Tourism | 60 | 3435 | 1131.7 |
| 25: Area Studies | 21 | 726 | 581.4 |
| 26: Modern Languages & Linguistics | 43 | 1565 | 1223.7 |
| 27: English Language & Literature | 91 | 1474 | 1351.6 |
| 28: History | 79 | 1964 | 1806.3 |
| 29: Classics | 17 | 227 | 198.4 |
| 30: Philosophy | 35 | 1036 | 919.7 |
| 31: Theology & Religious Studies | 30 | 302 | 283.6 |
| 32: Art & Design: History, Practice & Theory | 79 | 1714 | 1191.0 |
| 33: Music, Drama, Dance, Performing Arts, Film & Screen Studies | 76 | 948 | 809.3 |
| 34: Communication, Cultural & Media Studies, Library & Information Management | 55 | 1382 | 1056.9 |

If fractional counting is used rather than full counting, then the number of journal articles submitted to each UoA radically reduces in some and remains almost the same in others. The largest reduction is 90%, for Clinical Medicine. In general, the lower numbered UoAs have more collaborative research and greater reductions in the number of articles if fractional counting is used (Figure 2).

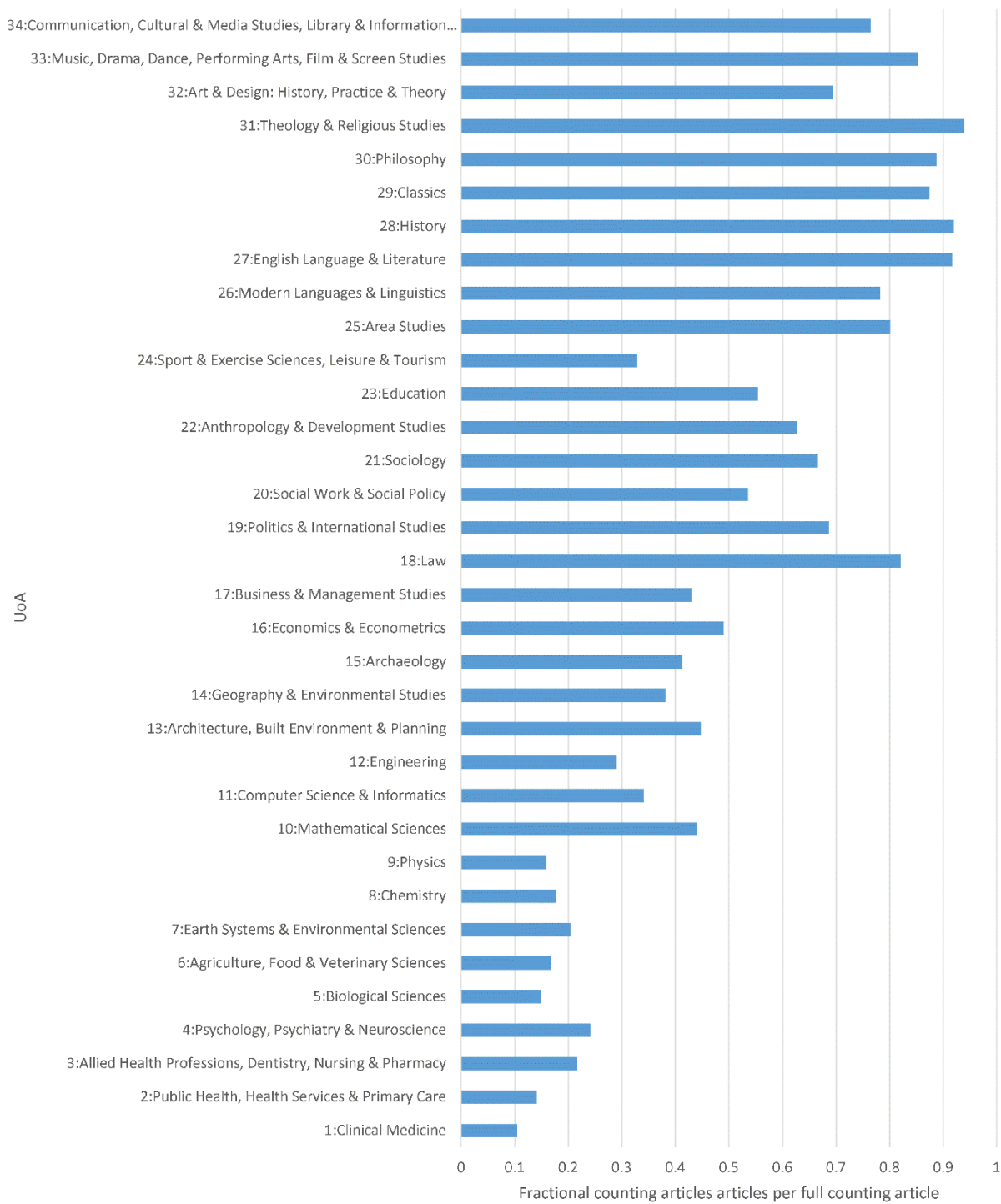

Figure 1. Number of fractional counting articles as a proportion of the number of full counting articles, by UoA.

If GPAs are calculated using fractional counting instead of the current system of full counting, then the relative scores for institutions change little within most UoAs (high correlations in Figure 2). The main exceptions are Archaeology (r=0.842) and Physics (r=0.884). Whilst these two correlations are high, they still reflect substantial changes between institutions due to fractional counting.

If the GPA advantage (GPA subtract author weighted GPA) is correlated against GPA then it is positive in most cases. This show that high scoring UoAs tend to gain from using full counting rather than fractional counting in most UoAs. The correlation is statistically significantly different from 0 (i.e., the 95% confidence interval excludes 0) in four cases: UoA 3 Allied Health Professions, Dentistry, Nursing & Pharmacy; UoA 15 Archaeology; UoA 20 Social Work & Social Policy; UoA 28 History; and UoA 31 Theology and Religious Studies. In these areas, higher scoring institutions had clearly gained from their collaborative articles having full weight.

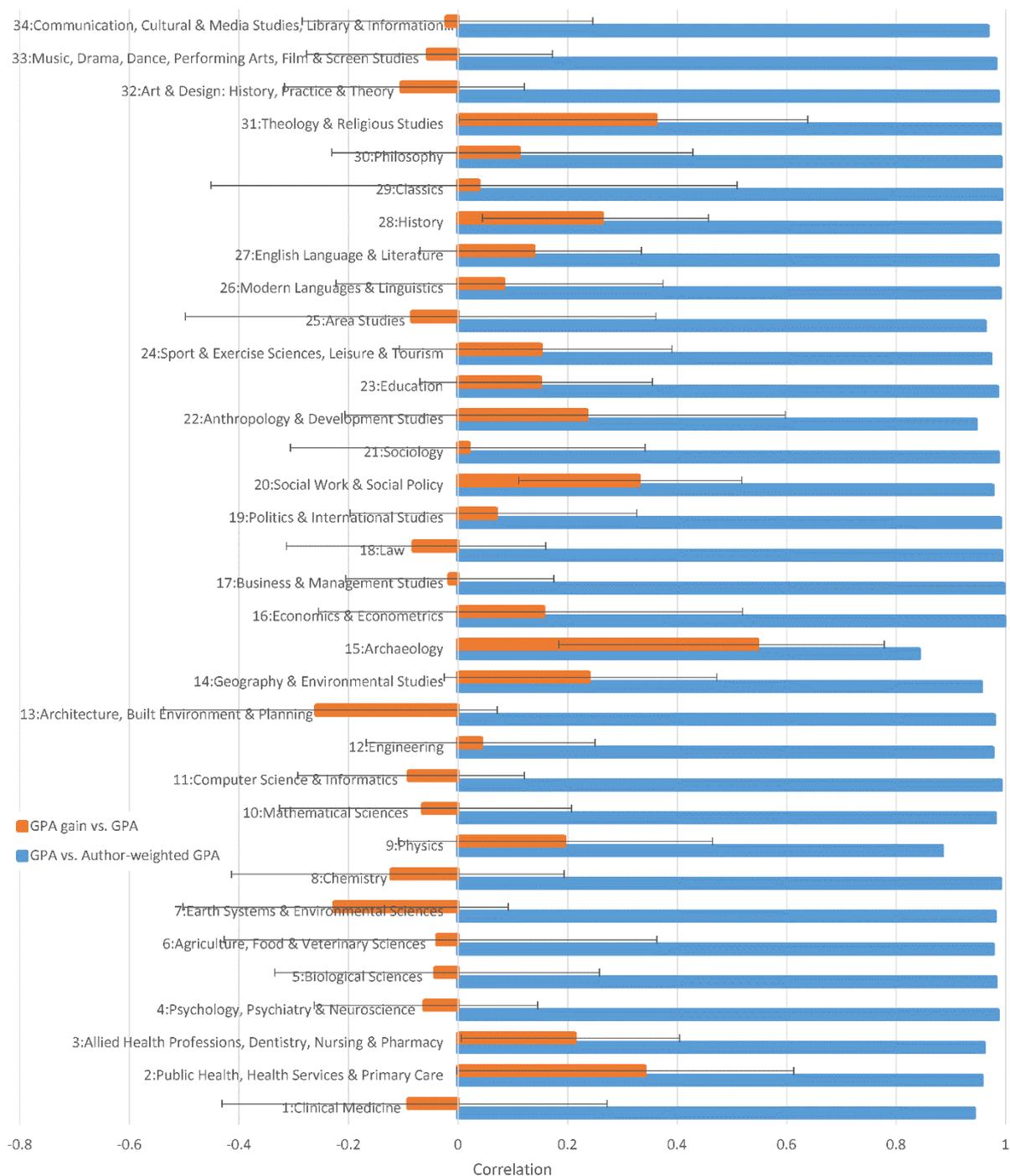

Figure 2. Institution-level Pearson correlations between institutional GPA and author-weighted GPA and between institutional GPA increase and institutional GPA. The GPAs include journal articles only. Error bars show 95% confidence intervals for the first correlation.

For UoA 15 Archaeology, with the largest GPA change between weighted and unweighted versions, there is an underlying reasonably linear trend between GPA and author weighted GPA (Figure 3). The diamonds furthest from the rest represent institutions with the largest GPA shift. For example, the institution with the second highest GPA has the 14th highest weighted GPA so it has benefitted substantially from full counting, at least for journal articles. This means that its more collaborative research was relatively high quality compared to its less collaborative or solo research. If the institution's best work went into this collaborative research, then a case could be made that the GPA is fair. On the other hand, if the institution's work was similar quality throughout and the quality of the most collaborative work was primarily boosted by the work of other team members then the GPA would be unfair. It is impossible to know which is true.

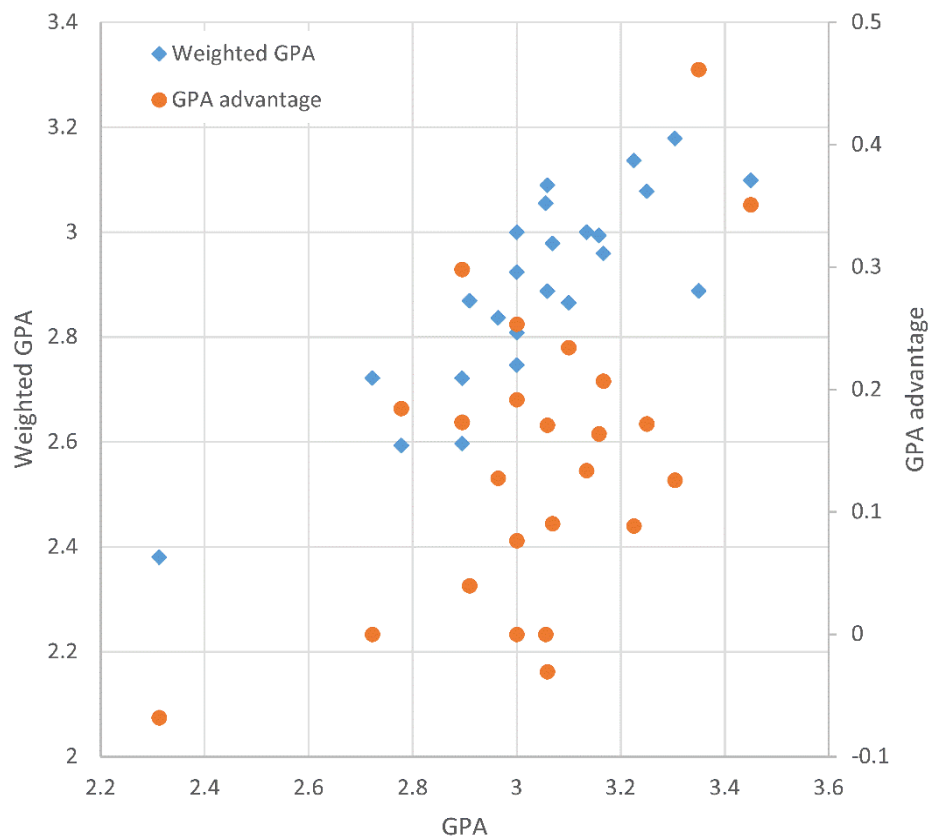

Figure 3. Author weighted GPA and GPA advantage against GPA for UoA 15 Archaeology.

For UoA 9 Physics, with the second largest GPA change between weighted and unweighted versions, there is again an underlying reasonably linear trend between GPA and author weighted GPA (Figure 4). The diamonds furthest from the rest represent again indicate institutions that have benefitted most from full counting. In this case, there is also one reverse outlier in the sense of a high GPA institution that benefitted the least from full counting. Physics includes areas with highly collaborative equipment consortia, such as high energy physics and astronomy, and lower collaboration areas, such as theoretical physics. It is therefore possible that the substantial score shifts were from departments with either different quality research specialisms in high and low collaboration areas. In this case, introducing fractional counting would work in favour of the quality of the low collaboration specialty.

High collaboration fields are sometimes also highly productive, which complicates the interpretation of the results. This is because each researcher can submit a maximum of 5 outputs to the REF. Thus, a large consortium researcher producing 100 papers per year in a huge team would only be able to submit 5 of them to the REF. Fractional counting would disadvantage such researchers' specialisms by effectively penalising them for splitting their work amongst too many papers to submit. The same is true to a much lesser degree for full counting.

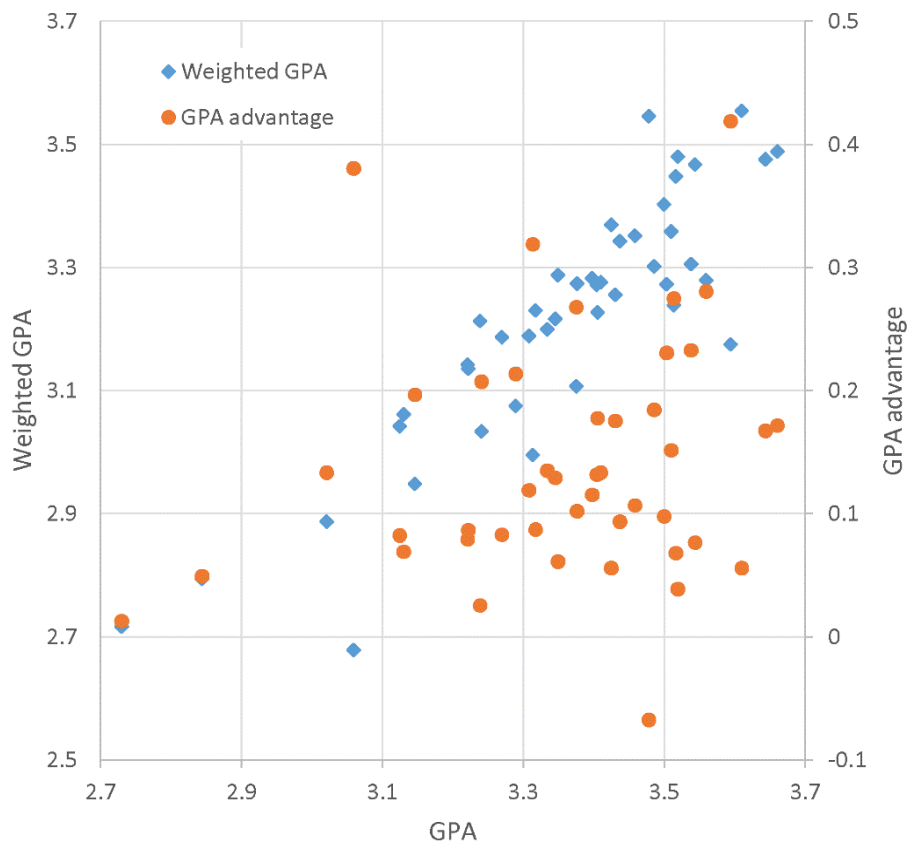

Figure 4. Author weighted GPA and GPA advantage against GPA for UoA 9 Physics.

## 4 Discussion and conclusions

The results should not be over-generalised because of several factors. First, the fractionalised counting method does not consider the proportion of authors from the submitting institution. For example, if all ten authors of a paper were from one university and only one author submitted it (multiple submissions from the same institution are normally not allowed), then the institution would get a 1/10 credit although a 10/10 credit would be fairer. Unfortunately, it was not possible to calculate the number of authors from the submitting institution for a paper because the REF considers the location of each author on the census date rather than their affiliation address, so many articles will have no authors with a submitting institution affiliation. Thus, both the whole and fractional counting methods are unfair: the first over-credits the submitting institution unless the paper is internal to the institution, and the second under-credits the submitting institution unless the author is the only institution member in the authorship team. Second, both methods probably inaccurately calculate the contribution of the submitting author in most non-solo papers. Third both methods ignore papers not submitted to the REF. It seems likely that researchers working in larger teams would write more papers and therefore submit a lower fraction of their outputs to the REF, but this penalises them in the fractionalised method.

The results suggest that switching from full counting to fractional counting when evaluating the average quality of journal articles from a department-level grouping tends to change the results only a small amount in some fields. The two main exceptions are archaeology and physics in REF2021. Because of the reasons given above it is impossible to deduce which ranking is better but the fact that the rankings are substantially different in two

UoAs shows that the decision has affected the reputations of some UK archaeology and physics departments. It is therefore a real concern rather than a purely theoretical issue.

These results apply to one relatively unusual research assessment system that limits the number of outputs per researcher and there may be different patterns for other field categorisation systems and countries, and for non-selective assessments. Nevertheless, it is the largest scale evidence so far that fractionalised counting usually has little effect on the average quality scores of departmental-level research groupings. Results for the exceptions (physics and archaeology) should be treated more carefully, however, and a special evaluation of the influence of collaboration is recommended for these. A corollary of the result is that in most fields, there is little evidence that institutions have gamed the system by somehow hijacking large collaborative projects to boost the average quality of their research.

## 5   Acknowledgement

This study was funded by Research England, Scottish Funding Council, Higher Education Funding Council for Wales, and Department for the Economy, Northern Ireland as part of the Future Research Assessment Programme (https://www.jisc.ac.uk/future-research-assessment-programme). The content is solely the responsibility of the authors and does not necessarily represent the official views of the funders.